\documentclass[twocolumn]{webofc}

\usepackage[varg]{txfonts}   
\usepackage{upgreek}

\newcommand{\be}{\begin{equation}}
\newcommand{\ee}{\end{equation}}
\newcommand{\ba}{\begin{align}}
\newcommand{\ea}{\end{align}}
\newcommand{\bea}{\begin{eqnarray}}
\newcommand{\eea}{\end{eqnarray}}

\begin{document}
\title{Hadronic contributions to the muon $g-2$ in holographic QCD}
%
%

\author{\firstname{Josef} \lastname{Leutgeb}\inst{1}\fnsep\thanks{\email{josef.leutgeb@tuwien.ac.at}} \and
        \firstname{Jonas} \lastname{Mager}\inst{1}\fnsep\thanks{\email{jonas.mager@tuwien.ac.at}} \and
        \firstname{Anton} \lastname{Rebhan}\inst{1}\fnsep\thanks{\email{anton.rebhan@tuwien.ac.at}}
}

\institute{Institut f\"ur Theoretische Physik, Technische Universit\"at Wien,
        Wiedner Hauptstrasse 8-10, A-1040 Vienna, Austria \label{addr1}
}

\abstract{%
  We discuss the recent progress made in using bottom-up holographic QCD models
  in calculating hadronic contributions to the anomalous magnetic moment of the muon, in particular the hadronic light-by-light scattering contribution, where holographic QCD naturally satisfies the Melnikov-Vainshtein constraint by an infinite series of axial vector meson contributions.
}
\maketitle
\fancyhead[RO,LE]{\thepage}

\section{Introduction}
\label{intro}

The anomaly of the magnetic moment of the muon $a_\upmu=(g-2)_\upmu/2$ \cite{Jegerlehner:2017gek} is currently known
with an experimental uncertainty of only $41\times 10^{-11}$
when the E821/BNL measurement from 2006 \cite{Muong-2:2006rrc} is
combined with the concordant result obtained in 2021 by
the Muon $g-2$ Collaboration at Fermilab
\cite{Muong-2:2021ojo}, corroborating a long-standing
discrepancy with the Standard Model prediction, which
according to the 2020
White Paper (WP) of the Muon $g-2$ Theory Initiative \cite{Aoyama:2020ynm}
has a similar estimated error of $43\times 10^{-11}$ but a deviation of
4.2 $\sigma$ in its value. The theoretical error is completely dominated
by hadronic contributions, which arise from hadronic vacuum polarization (HVP) and hadronic light-by-light (HLBL) scattering,
\bea
a_\upmu^{\text{HVP,WP}}=(6845 \pm 40)\times 10^{-11} \quad \text{(0.6\% error)},\nonumber\\
        a_\upmu^{\text{HLBL,WP}}=(92 \pm 19)\times 10^{-11} \quad \text{(20\% error}),
\eea
in (effective) one-loop and two-loop contributions in the muon-photon vertex diagram.

The WP result for the HVP contribution is obtained by a data-driven,
so-called dispersive approach, which however has been challenged by a recent
lattice QCD result with comparable errors but a central result that
is about 3\% larger \cite{Borsanyi:2020mff}, almost eliminating the discrepancy with the experimental result for $a_\upmu$, but instead
producing a $\sim3\sigma$ deviation from results based on the $R$-ratio in $e^+e^-\to\text{hadrons}$.

The WP estimate of the HLBL contribution is only partially determined by data
and involves input from hadronic models with larger uncertainties, 
albeit direct complete lattice evaluations
are now gradually bringing down their errors \cite{Blum:2019ugy,Chao:2021tvp}.

Holographic QCD (hQCD) is a conjectural approximation to strongly coupled
QCD (in its large-$N_c$ limit) based on the more well-established
AdS/CFT correspondence \cite{Maldacena:1997re,Aharony:1999ti} and experience with top-down string-theoretical
constructions of gauge/gravity dual theories \cite{Witten:1998zw,Sakai:2004cn,Sakai:2005yt}. Using a minimal set of parameters, holographic
QCD has proved to be capable of good qualitative and often quantitative
predictions in hadron physics \cite{Erlich:2005qh,DaRold:2005mxj,Hirn:2005nr} with typical errors of (sometimes less than) 10 to 30\%.
This is clearly too crude to be of help with the currently most
pressing issue of the theory result for $a_\upmu^\mathrm{HVP}$, where the discrepancy between lattice and data-driven approaches is a few percent and where eventually sub-percent accuracy is needed. However, in the case
of HLBL contributions, the error is dominated by two contributions,
the estimate of the effect of short-distance constraints (SDC) and
the contribution of axial-vector mesons, where the WP values are
\bea
a_\upmu^{\text{HLBL,SDC}}=(15 \pm 10)\times 10^{-11} \quad \text{(67\% error)},\nonumber\\
        a_\upmu^{\text{HLBL,axials}}=(6 \pm 6)\times 10^{-11} \qquad \text{(100\% error}).
\eea
It is here that hQCD can provide interesting results,
whereas the rather precisely known HVP contribution can be used
to assess potential deficiencies of the various hQCD models.

\section{Top-down and bottom-up hQCD models}

Exact holographic duals are known only in certain limits
for highly symmetric theories such as the superconformal $\mathcal{N}=4$ Yang-Mills theory in the limit of large-$N_c$ and infinite 't Hooft coupling $g^2N_c$. But
already in 1998, Witten \cite{Witten:1998zw} succeeded in constructing a gauge/gravity dual
to the low-energy limit of large-$N_c$ Yang-Mills theory, ``top-down'' from type-IIA superstring theory, where supersymmetry and conformal symmetry are
broken by compactification on one extra dimension beyond the
holographic dimension dual to inverse energy.
In 2004, Sakai and Sugimoto found a D-brane construction
within Witten's model that adds chiral quarks
in the fundamental representation \cite{Sakai:2004cn,Sakai:2005yt},
thus providing the so far closest dual theory to low-energy QCD
in the large-$N_c$ and chiral limit.
With having only a mass scale and one dimensionless number (the 't Hooft
coupling at the scale of the Kaluza-Klein compactification) the
(Witten-)Sakai-Sugimoto model turns out to be remarkably successful
both qualitatively and also quantitatively. Chiral symmetry breaking is a direct consequence of its geometry, while flavor anomalies are
naturally realized. However, above the
Kaluza-Klein scale $M_\mathrm{KK}\approx 1$ GeV this model exhibits
a different ultra-violet (UV) behavior than QCD, since its actual dual
at high energy scales is a 5-dimensional superconformal theory
instead of 4-dimensional QCD with asymptotic freedom.

Skipping a string-theoretic top-down derivation, phenomenologically
interesting ``bottom-up'' models of hadron physics were obtained in \cite{Erlich:2005qh,DaRold:2005mxj,Hirn:2005nr} that combine certain features also
present in the Sakai-Sugimoto model with a simpler geometry that is
asymptotically AdS$_5$ (and therefore conformal in the UV), breaking
conformal symmetry in the infrared (IR) by a hard-wall (HW) cutoff
with appropriate boundary conditions or by a soft wall (SW) provided by
a nontrivial dilaton \cite{Ghoroku:2005vt,Karch:2006pv,Kwee:2007dd,Gursoy:2007cb,Gursoy:2007er,Colangelo:2008us,Gherghetta:2009ac,Branz:2010ub,Colangelo:2011xk}, which has similarities with light-front holographic QCD \cite{Brodsky:2014yha}.

A common feature of the various bottom-up and also the top-down Sakai-Sugimoto model is that vector and axial vector mesons as well as pions
are described by 5-dimensional Yang-Mills fields
$\mathcal{B}^{L,R}_{M}=\mathcal{B}^{V}_{M}\mp \mathcal{B}^{A}_{M}$
for the global $U(N_f)_L\times U(N_f)_R$ chiral symmetry of boundary theory
with 5-dimensional action
 \bea
S_{\rm YM}
\propto
\frac{1}{g_5^2}
\;\text{tr}\int d^4x \int_0^{z_0} dz\,e^{-\Phi(z)}\sqrt{-g}\, g^{PR}g^{QS}
\nonumber\\
\times 
\left(\mathcal{F}^{(L)}_{PQ}\mathcal{F}^{(L)}_{RS}
+\mathcal{F}^{(R)}_{PQ}\mathcal{F}^{(R)}_{RS}\right),
\eea
where $P,Q,R,S=0,\dots,3,z$ and $\mathcal{F}_{MN}=\partial_M \mathcal{B}_N-\partial_N \mathcal{B}_M-i[\mathcal{B}_M,\mathcal{B}_N]$
with conformal boundary at $z=0$, and
either a sharp cut-off of AdS$_5$ at $z_0$ (the location of the hard wall) or $z_0=\infty$ (SW) with nontrivial dilaton.

Chiral symmetry is broken either by introducing an extra bifundamental
scalar field $X$ dual to quark bilinears $\bar q_L q_R$ as in the original HW model
of Erlich et al.~\cite{Erlich:2005qh,DaRold:2005mxj} (HW1), or
through different boundary conditions for vector and axial vector fields
at $z_0$ as in the Hirn-Sanz model \cite{Hirn:2005nr} as well as in
the top-down Sakai-Sugimoto model \cite{Sakai:2004cn,Sakai:2005yt}.

Vector meson dominance (VMD), in a form necessarily involving
an infinite tower of vector mesons, is naturally built in. Photons
are described by boundary values of the appropriate combination of
the vector gauge fields as those are sourced by quark currents,
and they couple through bulk-to-boundary propagators to mesonic
degrees of freedom described by normalizable modes.

Of particular importance for the following is that flavor anomalies follow uniquely
from 5-dimensional Chern-Simons terms $S_{\rm CS}^L-S_{\rm CS}^R$ with
\be\label{SCS}
            S_{\rm CS}=\frac{N_c}{24\pi^2}\int\text{tr}\left(\mathcal{B}\mathcal{F}^2-\frac{i}2 \mathcal{B}^3\mathcal{F}
            -\frac1{10}\mathcal{B}^5\right)
\ee
in differential-form notation.

\section{Anomalous TFFs from hQCD and HLBL contribution to the muon $g-2$}

\subsection{Holographic pion TFF}
\label{sec:TFFs}

\begin{figure}[t]
\medskip
\begin{center}
\includegraphics[width=0.275\textwidth,clip]{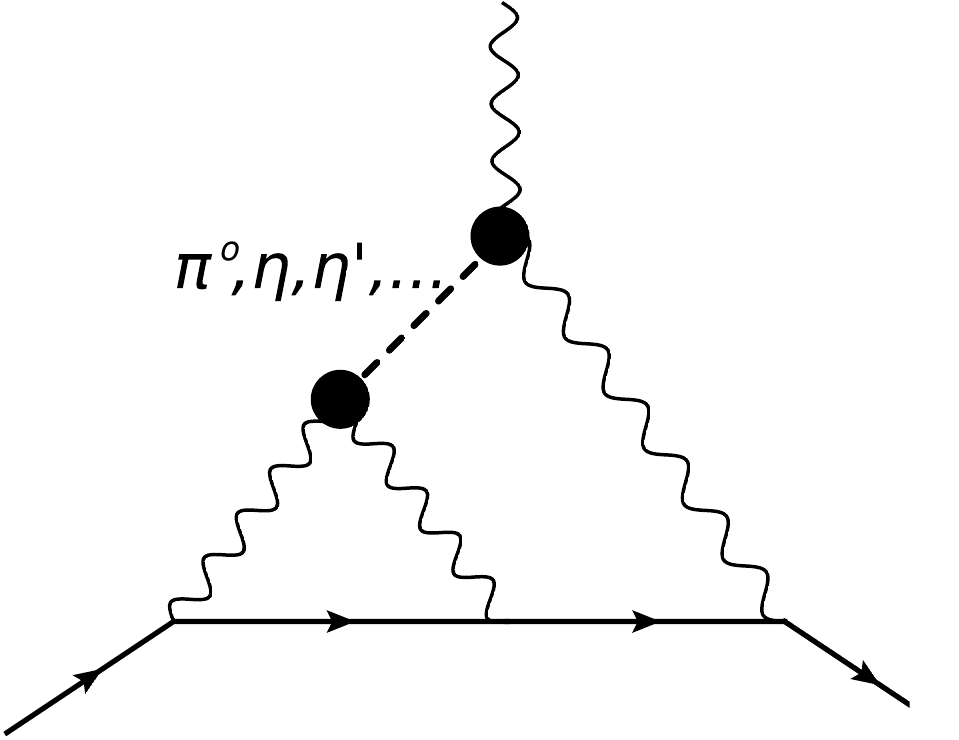}
\end{center}
\begin{picture}(0,0)
 \put(66,50){$Q_1$}
 \put(116,50){$Q_2$}
 \put(146,65){$Q_3$}
 \put(125,115){$Q_4=0$}
\end{picture}
\vspace*{-7mm}
\caption{Hadronic light-by-light scattering contribution to $a_\upmu$ from single meson exchange}
\label{fig:singlemesonexchange}       
\end{figure}
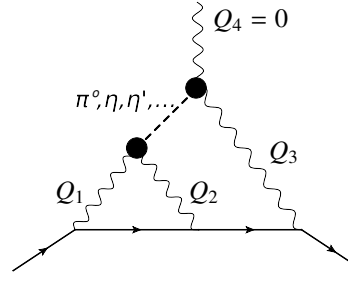

The leading HLBL contribution to $a_\upmu$ comes from the $\pi^0$ exchange diagram shown in fig.~\ref{fig:singlemesonexchange} due to
the anomalous coupling of the pion to two photons. This involves
both a singly-virtual transition form factor (TTF) in the upper vertex
and a doubly-virtual one in the interior of the diagram.
The holographic prediction is given by
\be
            F_{\pi^0\gamma^*\gamma^*}(Q_1^2,Q_2^2)=
            -\frac{N_c}{12\pi^2 f_\pi}\int_{0}^{z_{0}}dz\,\mathcal{J}(Q_1,z)\mathcal{J}(Q_2,z)
            \Psi(z),
\ee
where $\mathcal J(Q,z)$ is the bulk-to-boundary propagator of a photon
with virtuality $Q^2=-q^2$ and a holographic pion profile function $\Psi(s)$. This has been studied by Grigoryan and Radyushkin in \cite{Grigoryan:2007wn,Grigoryan:2008up,Grigoryan:2008cc}, who noticed that hQCD models
with asymptotic AdS$_5$ geometry reproduce the asymptotic momentum
dependence obtained by Brodsky and Lepage 
\cite{Brodsky:1981rp,Lepage:1979zb,Lepage:1980fj},
\bea
                &&F_{\pi^0\gamma^*\gamma^*}(Q_1^2,Q_2^2)\nonumber\\
                &&\to\frac{2 f_\pi}{Q^2}\left[ \frac1{w^2}-\frac{1-w^2}{2w^3}\ln\frac{1+w}{1-w} \right]
                \label{pionTFFas}
\eea
with $Q^2=\frac12(Q_1^2+Q_2^2)\to\infty$, $w=(Q_1^2-Q_2^2)/(Q_1^2+Q_2^2)$,
which is not achieved by conventional VMD models.

With certain simplifications the holographic results for the pion TFF
have been employed in ref.~\cite{Hong:2009zw,Cappiello:2010uy} to evaluate the $a_\upmu$ contribution given by the diagram in fig.~\ref{fig:singlemesonexchange}. In ref.~\cite{Leutgeb:2019zpq} we
have carried out a complete evaluation and found good agreement with the
data-driven (dispersive) result, which is bracketed by the HW1 and HW2 results when these models are fitted to $f_\pi$ and $m_\rho$. With such
a fit, the HW2 model, which has one parameter less than the chiral HW1 model, actually undershoots the asymptotic limit of \eqref{pionTFFas}
by 38\% (fitting the asymptotic limit would instead lead to an overweight $\rho$ meson with mass of 987 MeV). Its prediction of $a_\upmu^{\pi^0}=56.9\times 10^{-11}$ is correspondingly on the low side. The HW1 model,
which satisfies \eqref{pionTFFas} exactly, predicts $a_\upmu^{\pi^0}=65.2\times 10^{-11}$, which is consistent with, but somewhat larger than, the WP value $62.6^{+3.0}_{-2.5}\times 10^{-11}$.

However, at not too large, phenomenologically relevant energy scales, the asymptotic behavior of the TFF is modified by gluonic corrections of the order of 10\% \cite{Melic:2002ij,Bijnens:2021jqo}, while hQCD models with a simple AdS$_5$ background approach the asymptotic limit somewhat too quickly. Similar corrections but with opposite sign appear in the vector correlator appearing in HVP. Indeed, the HW1 model with complete UV fit underestimates HVP. Reducing $g_5^2$ by 10\% (which is achieved by
fitting instead the decay constant of the $\rho$ meson) brings 
the HVP in line with the dispersive result (to within 5\%) \cite{Leutgeb:2022cvg} and also makes the HLBL result $a_\upmu^{\pi^0}$ completely 
coincident with the central WP value \cite{Leutgeb:2021mpu}.

\begin{figure}[t]
\centerline{\small $Q^2 F_{P\gamma^*\gamma}(Q^2,0)$ [GeV]\hfill}
\centerline{\includegraphics[width=0.4\textwidth,clip]{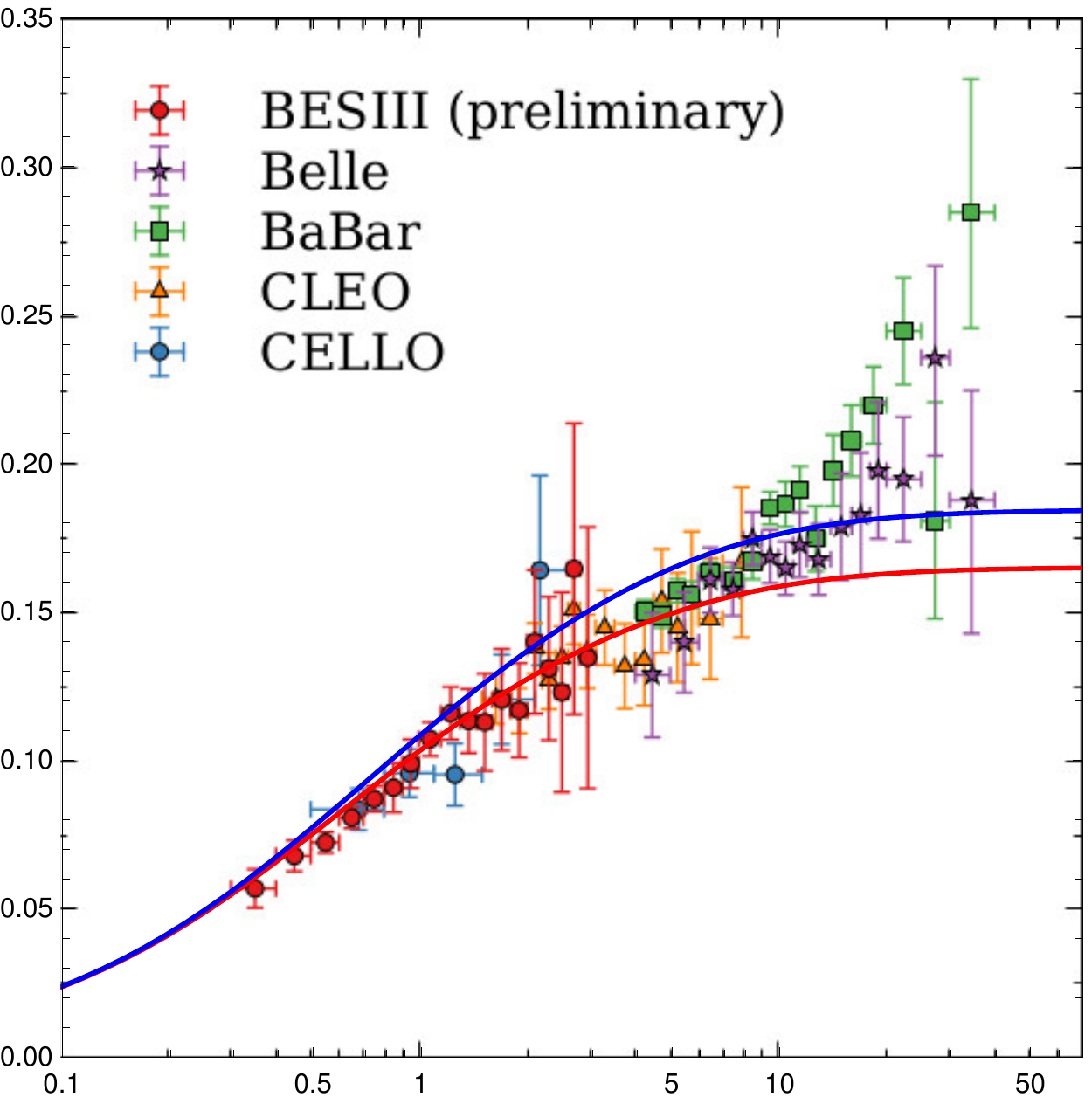}}
\centerline{\small $Q^2$ [GeV$^2$]}
\caption{Holographic results for the single virtual TFF $Q^2 F(Q^2,0)$ for $\pi^0$, plotted on top of experimental data as compiled in Fig.~53 of Ref.~\cite{Aoyama:2020ynm}
for $g_5=2\pi$ (OPE fit, blue) and the reduced value (red)
corresponding to a fit of $F_\rho$. (Taken from \cite{Leutgeb:2022lqw})}
\label{fig:pionTFFsv}
\end{figure}

\begin{figure}[t]
\centering
\includegraphics[width=0.46\textwidth,clip]{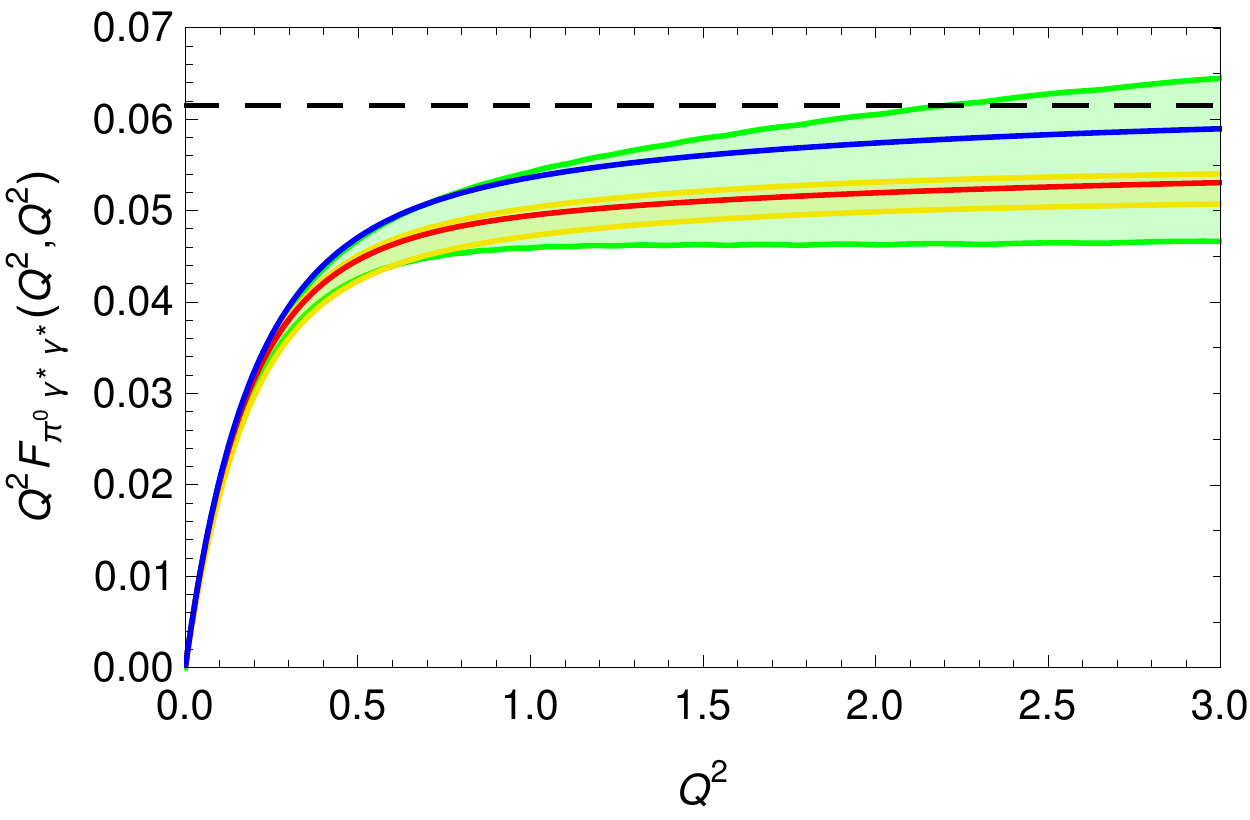}
\caption{Holographic results for the doubly virtual $F_{\pi^0\gamma^*\gamma^*}$ compared to the dispersive result of Ref.~\cite{Hoferichter:2018kwz} (green band) and the lattice result of Ref.~\cite{Gerardin:2019vio} (yellow band); the OPE limit is indicated by the dashed horizontal line. The blue line corresponds to $g_5=2\pi$ (OPE fit) and the red one to the reduced value $g_5$ ($F_\rho$-fit). (Taken from \cite{Leutgeb:2022lqw})}
\label{fig:pionTFFdv}
\end{figure}

In fig.~\ref{fig:pionTFFsv} the HW1 result (with finite quark masses) for the singly virtual pion TFF is compared with experimental data, where the
result with reduced $g_5^2$ is seen to agree somewhat better at the relevant energy scale $Q^2\lesssim 10\,\mathrm{GeV}^2$; in fig.~\ref{fig:pionTFFdv} the doubly virtual result for $Q_1^2=Q_2^2$ is
compared with the dispersive result of ref.~\cite{Hoferichter:2018kwz} and the lattice result of Ref.~\cite{Gerardin:2019vio}. Here the HW1 result with reduced coupling coincides with the central result of the dispersive approach within line thickness and it also happens to be in the center of the lattice result.

\subsection{Axial vector meson TFF}

In hQCD, the coupling of axial vector mesons to photons is determined
by the same action \eqref{SCS} that gives rise to the anomalous TFFs of
the pseudoscalars. There is an infinite tower of axial vector mesons and their decay amplitude has the form \cite{Leutgeb:2019gbz,Cappiello:2019hwh}
\bea\label{calMa}
            &&\mathcal{M}_{\mathcal{A}\gamma^*\gamma^*}=i\frac{N_c}{4\pi^2}\mathrm{tr}(\mathcal{Q}^2 t^a)\,\epsilon_{(1)}^\mu \epsilon_{(2)}^\nu
            \epsilon_\mathcal{A}^{*\rho} \epsilon_{\mu\nu\rho\sigma}\nonumber\\
            &&\;\times\left[q_{(2)}^\sigma Q_1^2 A_n^a(Q_1^2,Q_2^2)-q_{(1)}^\sigma Q_2^2 A_n^a(Q_2^2,Q_1^2)\right],
\eea
involving an asymmetric structure function
        \be
            A_n(Q_1^2,Q_2^2) = \frac{2g_5}{Q_1^2} \int_0^{z_0} dz \left[ \frac{d}{dz} \mathcal{J}(Q_1,z) \right]
            \mathcal{J}(Q_2,z) \,\psi^A_n(z) 
        \ee
where $\psi^A_n(z)$ is the holographic wave function of the $n$-th axial vector meson. The most general decay amplitude of axial vector mesons would permit a second asymmetric structure function \cite{Pascalutsa:2012pr,Roig:2019reh,Zanke:2021wiq}, but this does not appear in the simplest versions of hQCD considered here.

The holographic result satisfies the Landau-Yang theorem \cite{Landau:1948kw,Yang:1950rg}, which forbids the decay of an axial vector meson in two real photons, due to the fact that
$\mathcal{J}'(Q,z)=0$ for $Q^2=0$.

The asymptotic form of $A_n(Q_1^2,Q_2^2)$ for large virtualities is
given by \cite{Leutgeb:2019gbz}
\begin{align}
                &A_n(Q_1^2,Q_2^2) \to \nonumber\\
                &\frac{12\pi^2 F^A_{n}}{N_c Q^4}
                \frac1{w^4}\left[
                w(3-2w)+\frac12 (w+3)(1-w)\ln\frac{1-w}{1+w}
                \right],
\end{align}
$w=(Q_1^2-Q_2^2)/(Q_1^2+Q_2^2)$,
which agrees with the independent pQCD result obtained recently in \cite{Hoferichter:2020lap}.

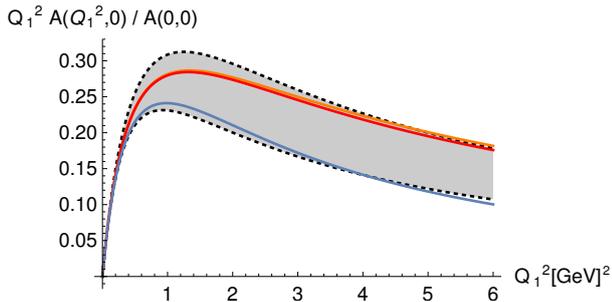
\begin{figure}[t]
\centering
\includegraphics[width=0.46\textwidth,clip]{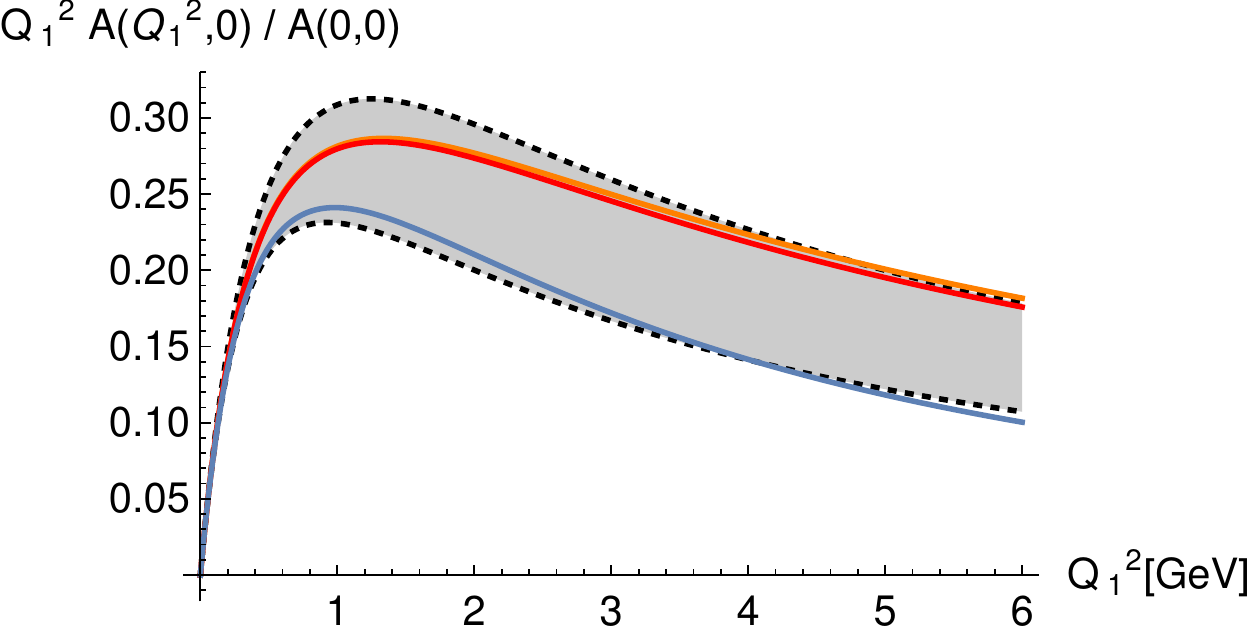}
\begin{picture}(0,0)
    \put(-2,12){\tiny\sf 2}
\end{picture}
\caption{Single-virtual axial vector TFF from holographic models 
(SS: blue, HW1: orange, HW2: red) compared with
dipole fit of L3 data for $f_1(1285)$ (grey band). The parameters
of all models are fixed by matching $f_\pi$ and $m_\rho$.
The results for HW1 and HW2 almost coincide, with HW2 at most a line
thickness above HW1. (Taken from \cite{Leutgeb:2019gbz})}
\label{fig:xf1ff}
\end{figure}

\begin{figure}[t]
\centering
\includegraphics[width=0.46\textwidth,clip]{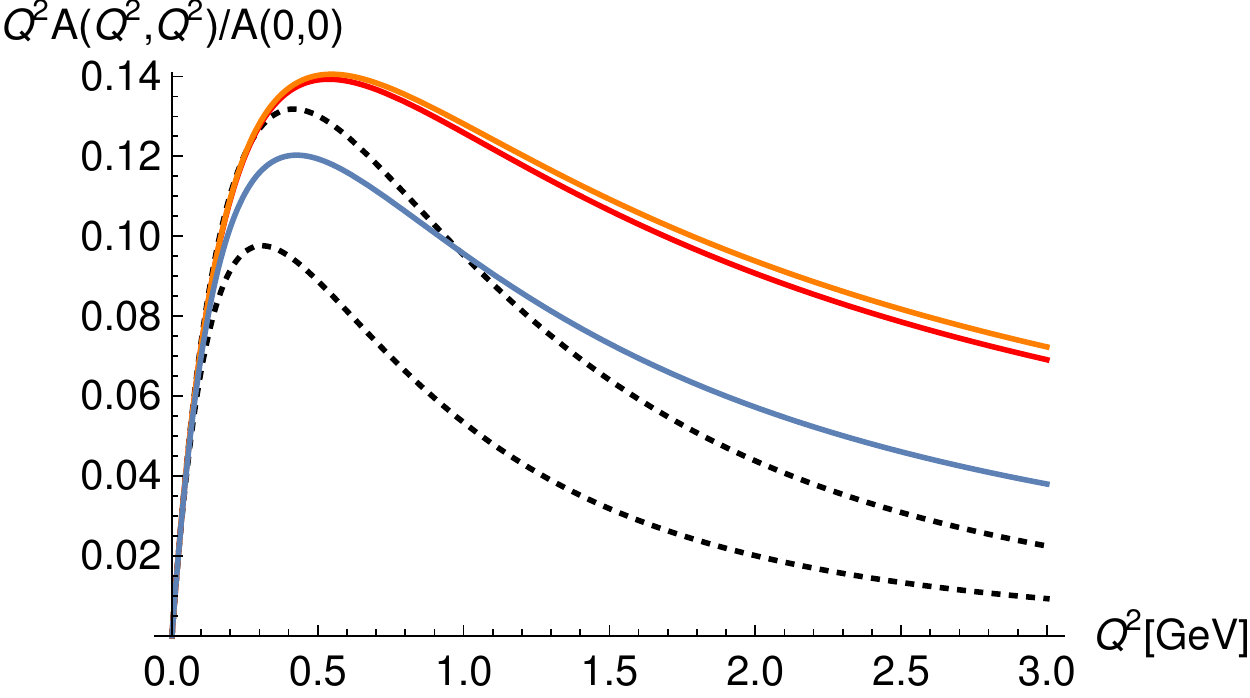}
\begin{picture}(0,0)
    \put(-2,12){\tiny\sf 2}
\end{picture}
\caption{Double-virtual axial vector TFF for $Q_1^2=Q_2^2=Q^2$ from holographic models 
(SS: blue, HW1: orange, HW2: red). The black dashed lines denote the extrapolation
of L3 data with a dipole model for each virtuality as used in the calculation of $a_\upmu^{f_1}$ in Ref.~\cite{Pauk:2014rta}. (Taken from \cite{Leutgeb:2019gbz})}
\label{fig:xf1ffsym}
\end{figure}
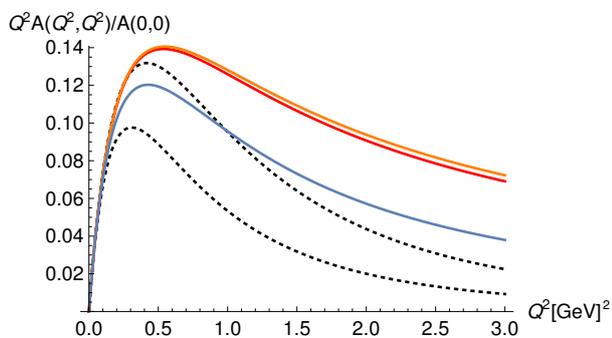

At moderate virtualities, the shape of the singly virtual axial TFFs
turns out to be consistent with the dipole fit of data obtained by the L3
experiment for $f_1(1285)$ \cite{Achard:2001uu}, as shown in fig.~\ref{fig:xf1ff}. The overall magnitude, which for HW1 and HW2 models
is found as $21.04$ and $16.63$ GeV$^{-2}$ is in roughly the right
ballpark compared to experimental data, albeit somewhat too high for HW1
which overestimates the equivalent photon widths of $f_1(1285)$ and $f_1'=f_1(1420)$ \cite{Achard:2001uu,Achard:2007hm}, but consistent with
the phenomenological value $A(0,0)_{a_1(1230)}=19.3(5.0) \,\mathrm{GeV}^{-2}$ obtained in \cite{Roig:2019reh,Masjuan:2020jsf}.

The results from the L3 experiment have been used by Pauk and Vanderhaeghen \cite{Pauk:2014rta} 
to estimate the contributions of $f_1$ and $f_1'$ to $a_\upmu$, assuming
a symmetric factorized dipole moment ${A^{\mathrm{PV}}(Q_1^2,Q_2^2)}/{A(0,0)}=(1+Q_1^2/\Lambda_D^2)^{-2}(1+Q_2^2/\Lambda_D^2)^{-2}$.
This drops too quickly in the doubly virtual case to match the short-distance behavior required by pQCD. In fig.~\ref{fig:xf1ffsym} the hQCD results are shown and compared with this simple ansatz, exhibiting
a considerable difference.

\subsection{Melnikov-Vainshtein SDC}

In \cite{Leutgeb:2019gbz,Cappiello:2019hwh} it has been shown moreover 
that in hQCD the infinite tower of axial vector mesons is responsible for the correct implementation of the Melnikov-Vainshtein SDC \cite{Melnikov:2003xd} for the HLBL scattering amplitude, which is a consequence of the nonrenormalization theorem for the chiral anomaly. In terms of the tensor basis used in \cite{Colangelo:2015ama} it reads
\begin{equation}
\label{eq:MVConstr}
    \lim_{Q_3\rightarrow \infty}\lim_{Q \rightarrow \infty}  Q_3^2 Q^2 \bar{\Pi}_1(Q,Q,Q_3)= -\frac{2}{3 \pi^2}.
\end{equation}
Clearly, each single meson exchange contribution gives a vanishing
result in this limit, because the propagator in fig.~\ref{fig:singlemesonexchange} goes like $1/Q_3^2$ and the two
form factors go like $1/Q^2$ and $1/Q_3^2$.

\begin{figure}[t]
\centerline{\includegraphics[width=0.6\textwidth,clip]{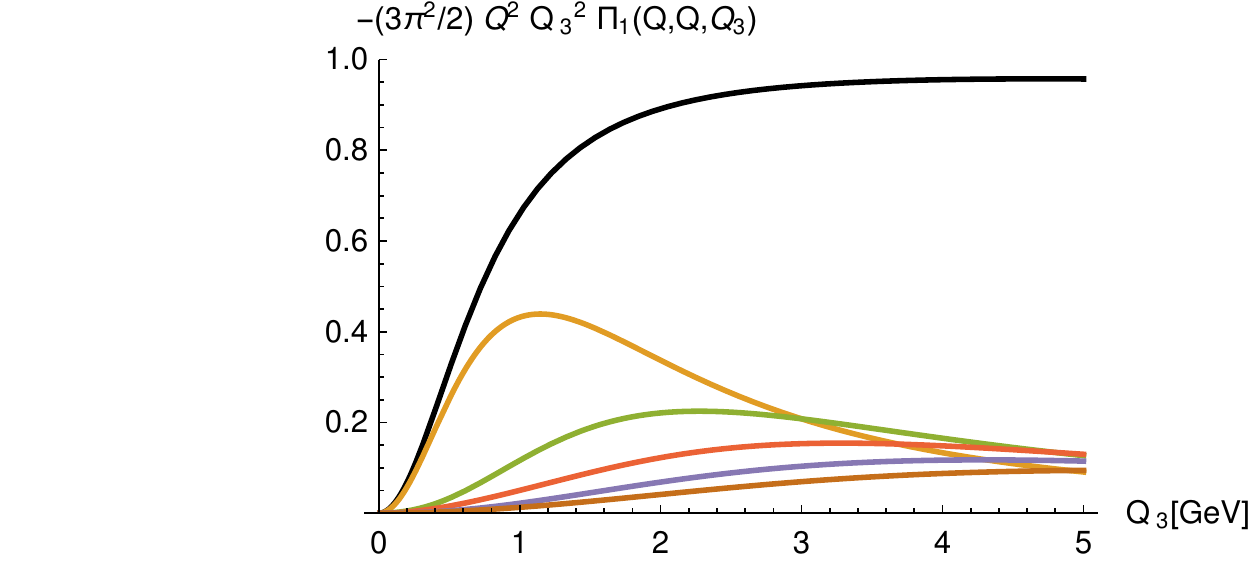}\qquad\qquad\qquad}
\caption{Axial-vector contribution to
$Q_3^2 Q^2 \bar\Pi_1(Q,Q,Q_3)$ as a function of $Q_3$ at $Q=50$ GeV
in the HW2 model (with $g_5=2\pi$ so that the SDC is satisfied to 100\%). 
The black line corresponds to the infinite sum over the tower of axial vector mesons, and the colored lines
give the contributions of the 1st to 5th lightest axial vector mesons. (Taken from \cite{Leutgeb:2019gbz})
}
\label{fig:MVlimit}
\end{figure}

In \cite{Melnikov:2003xd} Melnikov and Vainshtein proposed to estimate the impact of the MV-SDC by replacing the external TFF by its constant
on-shell value. With current input data, this would lead to an increase of almost 40\% of the pseudoscalar contribution $a_\upmu^{\pi^0,\eta,\eta'}$ by $38\times 10^{-11}$. However, the MV-SDC can also be satisfied by having an infinite tower of single-meson contributions. In Ref.~\cite{Colangelo:2019lpu,Colangelo:2019uex} Colangelo et al.\ showed that a Regge model for a tower of excited pseudoscalars can be constructed that saturates the MV-SDC with a contribution $\Delta a_\upmu^\mathrm{PS}=13(6)\times 10^{-11}$ when using the available experimental constraints for the photon decay amplitudes of excited pseudoscalars. However, as also noted in \cite{Colangelo:2019uex}, excited pseudoscalar states decouple in the chiral large-$N_c$ limit,
which is not the case for the infinite tower of axial vector mesons.
Indeed, in holographic QCD it is the latter which are responsible
for the implementation of the MV-SDC \cite{Leutgeb:2019gbz,Cappiello:2019hwh}. This is illustrated in fig.~\ref{fig:MVlimit} for the simple HW2 model, which permits a closed form solution for the full series, for $\Pi_1(Q,Q,Q_3)$
with large $Q=50$ GeV and increasing $Q_3\ll Q$: while each axial vector meson mode gives an asymptotically vanishing contribution, their infinite series satisfies the MV-SDC constraint.

In \cite{Leutgeb:2021mpu} two of us have shown that also in the HW1 model with massive quarks the infinite tower of axial vector mesons is responsible for the realization of the MV-SDC. The infinite tower of excited pseudoscalars, which does not decouple from the anomaly relation away from the chiral limit, contributes only subleading terms $\propto \ln(Q_3^2)/Q_3^4 Q^2$ to \eqref{eq:MVConstr}.

The hQCD models generally satisfy the MV-SDC to the same level as they
satisfy the SDCs on the TFFs (see the remarks in sect.\ \ref{sec:TFFs}.)
In the limit where all $Q_i$ become large simultaneously, another SDC
can be derived \cite{Melnikov:2003xd,Bijnens:2019ghy}
for $\bar\Pi$, $\lim_{Q\to\infty}Q^4\bar\Pi_1(Q,Q,Q)=-4/9\pi^2$.
In hQCD models with correct short-distance limit of the TFFs, the coefficient
on the right-hand side comes out 19\% smaller \cite{Cappiello:2019hwh,Leutgeb:2021mpu}.

\subsection{$a_\upmu$ contributions}

\begin{figure}[t]
\centering
\includegraphics[width=0.47\textwidth,clip]{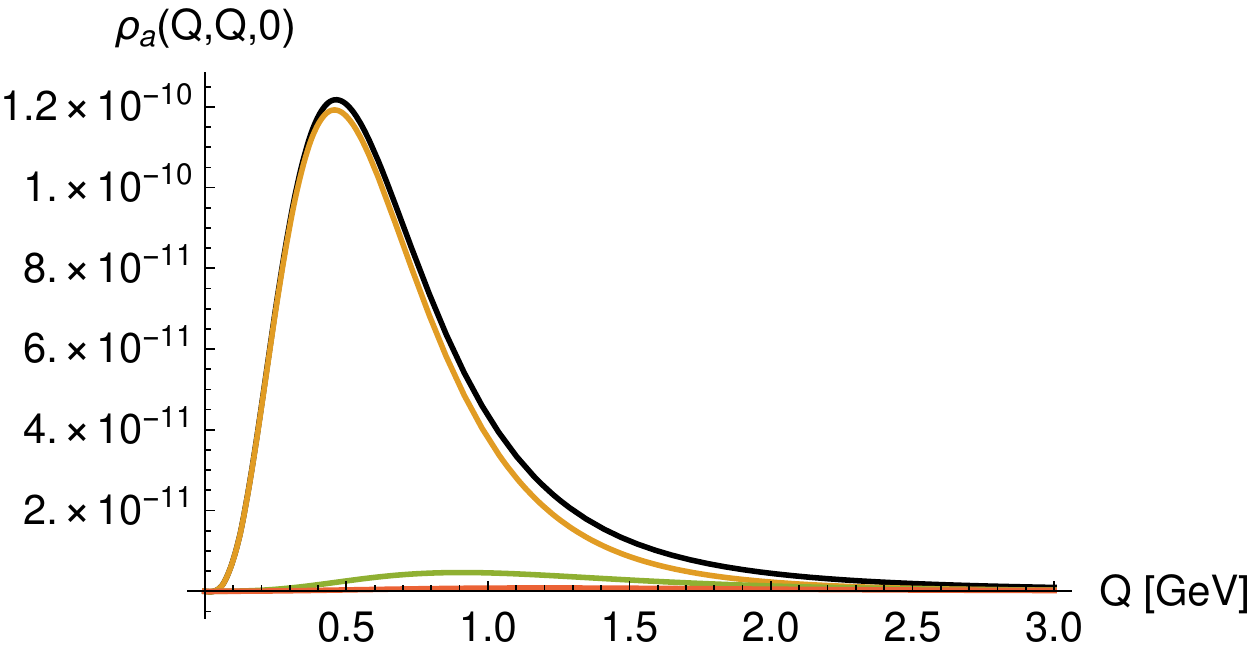}
\caption{The integrand $\rho_a(Q_1,Q_2,\tau)$ in (\ref{amuAVint}) in units of GeV$^{-2}$
for $Q_1=Q_2=Q$ and $\tau=0$ in 
the case of the HW2 model. The black line is
the result from the infinite sum over the tower of axial vector mesons, the colored lines
give the contributions of the 1st to 3rd lightest axial vector meson multiplets.
 (Taken from \cite{Leutgeb:2019gbz})
}
\label{fig:rhoaAV}
\end{figure}

Fig.~\ref{fig:rhoaAV} shows the individual contributions of
axial vector meson excitations to the integrand in
\be\label{amuAVint}
a_\upmu^\mathrm{AV}=\int_0^\infty dQ_1 \int_0^\infty dQ_2 \int_{-1}^1 d\tau \,\rho_a(Q_1,Q_2,\tau)
\ee
for $Q_1=Q_2$ and $\tau=0$. The ground-state axial vector meson mode clearly
dominates, but the infinite sum of excited modes does contribute non-negligibly.
For chiral models, this is shown quantitatively in table \ref{tab:total}.
In the Sakai-Sugimoto model, where the SDCs are not satisfied at all, excited modes contribute only a few percent.
In the HW2 model, which satisfies the SDCs qualitatively, but quantitatively only at the level of 62\% when
$f_\pi$ and $m_\rho$ are fitted to physical values, the excited modes
add +25\% to the contribution of the ground-state mode ($j=1$). 
This percentage increases to about 30\% in the HW1 model, where
the SDCs are matched exactly. Also shown in table \ref{tab:total}
are two modified versions of the HW1 model with $g_5^2$ reduced by 10\% and 15\%, which can be viewed as taking into account next-to-leading
order corrections at large but still physically relevant energy scales.
Also in these models, the excited modes add around 30\%, if not more, to $a_\upmu^{\mathrm{AV}}({j=1})$.

\begin{table*}
\centering
\caption{Axial vector contributions in chiral HW models with different levels of saturation of SDCs. In the chiral, flavor U(3)-symmetric case we have simply $a_\upmu^{\mathrm{AV}}=a_\upmu^{a_1+f_1+f_1'}=4 a_\upmu^{a_1}$. Also given are the masses $m_{a_1}$ and the photon coupling amplitudes $A_1(0,0)$ of the ground-state axial vector mesons ($j=1$).}
\label{tab-1}       
\begin{tabular}{llcccccccc}
\hline
 chiral model    &$m_{a_1}$ & $A_1(0,0)$ & SDC & $a_\upmu^{\mathrm{AV}}(j=1)$ & $(j\le2)$ & $(j\le3)$ & $(j\le4)$ & $(j\le5)$ & $a_\upmu^\mathrm{AV}$(all)\phantom{$^{1^{1^{1^1}}}$} \\
    \hline\\[-8pt]
    HW1 &1.3755 & 21.04 & 100\% & 31.4 & 36.2 & 37.9 & 39.1 & 39.6 & 41 $\times 10^{-11}$ \\
    HW1- &1.295 & 19.65 & 90\% & 28.7 & 34.3 & 36.0 & 37.1 & 37.5 & 38 $\times 10^{-11}$ \\
    HW1-{}- &1.255 &18.90 & 85\% & 27.3 & 33.2 & 34.9 & 35.9 & 36.3 & 37 $\times 10^{-11}$ \\
    HW2 & 1.235 & 16.63 & 62\% & 23.0 & 26.2 & 27.4 & 27.9 & 28.2 & 29 $\times 10^{-11}$ \\
    SS & 1.1865 & 15.93 & 0\% & 13.8 & 14.5 & 14.7 & 14.8 & 14.8 & $ 15 \times 10^{-11}$ \\
    \hline
    \end{tabular}
\end{table*}

While being completely absent in the chiral HW2 model, excited pseudoscalars, which decouple from the axial current in the chiral limit, always contribute to $a_\upmu$ in the HW1 model and its variants.
This has been studied in \cite{Leutgeb:2021mpu} in the flavor-symmetric
case with and without modified scaling dimension of the bifundamental scalar and also for different boundary conditions on the hard wall.
The various models show some differences in the composition of the
contributions to $a_\upmu$, but hardly deviate from the sum total in
the chiral model (where the physical pion mass is inserted manually
in the pion propagator). 

\begin{figure}[h]
\centering
\includegraphics[width=0.45\textwidth,clip]{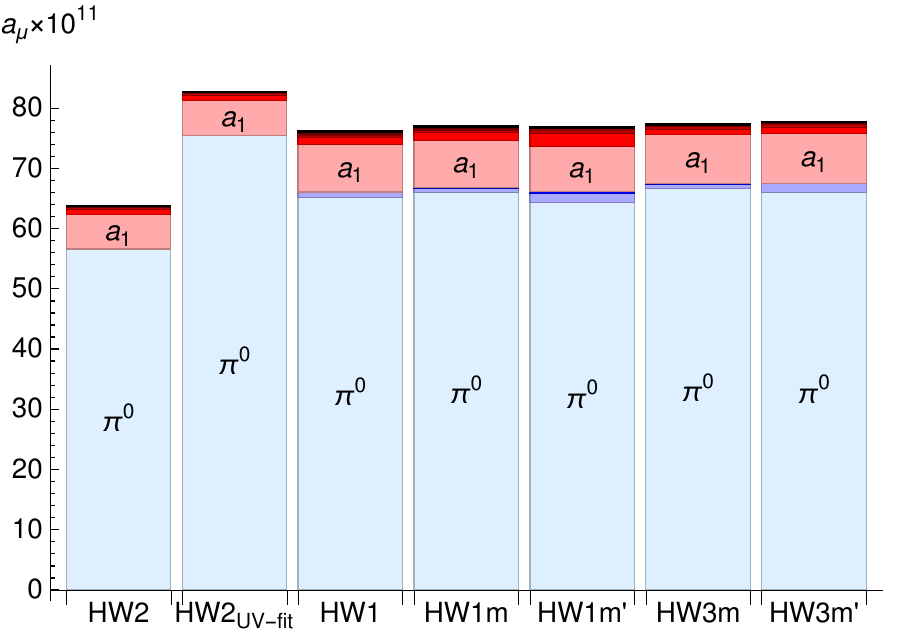}
\caption{Bar chart of the individual contributions to $a_\upmu$ from the isotriplet sector
in the Hirn-Sanz model fitted to IR (HW2) and UV data (HW2$_\text{UV-fit}$), the chiral HW1 model and various extensions to the massive case, with excited modes given by increasingly darker colors, blue for the $\pi^0$'s, red for the $a_1$'s. (Taken from \cite{Leutgeb:2021mpu})}
\label{fig:barchart}
\end{figure}

In the flavor-U(3)-symmetric case the total axial vector sector is four times the contribution of the $a_1$ meson. With a 10\% reduction of $g_5^2$ and
correspondingly the level of saturation of the SDCs, the massive HW models together yield 
$a_\upmu^\mathrm{AV}=4a_\upmu^\mathrm{a_1}={39.5(1.6)}\times 10^{-11}$, where approximately 58\% are due to longitudinal contributions entering $\bar\Pi_1$.
The contribution of excited pseudoscalars varies somewhat more, as can be seen from fig.~\ref{fig:barchart}. In models where their contribution is larger, the ground-state yields less, but we have
$a_\upmu^\mathrm{PS^*}\gtrsim 3.3\times 10^{-11}$ (again with U(3) symmetry and with 10\% reduction of $g_5^2$).
Even without the latter contribution, the holographic result is somewhat higher than the WP estimate for the combined effect of
axial vector mesons and SDCs, $a_\upmu^\mathrm{axials+SDC(WP)}=21(16)\times 10^{-11}$.

\subsection{Katz-Schwartz model: HW1 with solved U(1)$_A$ problem}

In the above estimates
obtained from flavor-U(3)-symmetric HW models, the 
effects of the large strange quark mass and the anomalous breaking of U(1)$_A$ symmetry has been ignored (in \cite{Leutgeb:2019gbz,Cappiello:2019hwh} the contribution of $\eta$ and $\eta'$ mesons, where U(3) symmetry clearly is no good approximation, has been estimated by
manually setting their masses and decay constants to physical values).

In \cite{Leutgeb:2022lqw} we have recently considered the minimal
extension\footnote{In \cite{Cappiello:2021vzi} an extension of the HW1 model was considered which in contrast to the minimal models leads to nonvanishing HLBL contributions of scalar mesons. While results were obtained that are consistent with previous estimates of these contributions, the scalar TFFs obtained therein do not agree with pQCD SDC constraints \cite{Hoferichter:2020lap}.} of the HW1 proposed by Katz and Schwartz \cite{Katz:2007tf,Schafer:2007qy}
to implement the U(1)$_A$ anomaly through an additional complex scalar $Y$ whose magnitude and phase are dual to the gluon operators $\alpha_s G_{\mu\nu}^2$ and $\alpha_s G\tilde G$. There a coupling term
$\kappa Y^{N_f} \det(X) \subset \mathcal{L}$ in the 5-dimensional theory
accounts for U(1)$_A$ anomalous Ward identities. It turns out that all results are weakly dependent on the new coupling $\kappa$ when $\kappa\gg1$. The only new free parameter is the value of the gluon condensate $\Xi$ which can be prescribed in the choice of the background
field $\langle Y\rangle=C+\Xi z^4$, while $C$ is fixed by OPE and the U(1)$_A$ anomaly to $C\propto \alpha_s$. The latter is modeled by
$\alpha_s\to 1/\beta_0 \ln(\Lambda_{QCD}z)$, $\Lambda_{QCD}\to z_0^{-1}$.
This model implements a Witten-Veneziano mechanism for the mass of isosinglet pseudoscalars, which are augmented by a pseudoscalar glueball mode mixing with the $\eta$ and $\eta'$ mesons.

In \cite{Katz:2007tf}, Katz and Schwartz have found that their model
can account for the masses of $\eta$ and $\eta'$ mesons with deviations of about $10\%$. By turning on the gluon condensate parameter $\Xi$, we have found \cite{Leutgeb:2022lqw} that this can be improved to the percent level. 

In the case of axial vector mesons, this model is somewhat less successful: the masses of $f_1$ and $f_1'$ are raised too much compared to their experimental values, and also the experimental values of their mixing is not reproduced. Nevertheless, this provides a first glimpse of what a more complete hQCD model can give. 

The results that we have obtained
with the Katz-Schwartz model augmented by a gluon condensate are given
in table \ref{tab:total}. In the isotriplet (pion and $a_1$) sector, the results are essentially unchanged, with the pion contribution matching perfectly the dispersive result, in particular when $g_5$ is fitted to match $F_\rho$, corresponding to a 90\% level of saturation of the OPE constraints. The contributions from $\eta$ and $\eta'$ mesons are then also completely in line with the WP estimates. Excited pseudoscalars (including a pseudoscalar glueball mode) contribute another $1.6\times 10^{-11}$, so that the total contribution of pseudoscalars is at the upper bound of the WP estimate. The combined contribution of ground-state axial vector mesons equals around 3.5 times the $a_1$ contribution, which is a moderate reduction of the factor 4 in the flavor-U(3) symmetric case. Somewhat surprisingly, the contribution of excited $f_1$ and $f_1'$ axial vector mesons is more strongly reduced, so that the ``best guess''
for the axials + LSDC contributions in this model turns out to be
$a_\upmu^\mathrm{AV+LSDC}=30.5\times 10^{-11}$, and
$a_\upmu^\mathrm{AV+PS^*+LSDC}=32\times 10^{-11}$ when excited pseudoscalars are included.
Both results are somewhat above but now safely within the estimated error of the WP
estimate $a_\upmu^\mathrm{axials+SDC(WP)}=21(16)\times 10^{-11}$.

We should like to point out, however, that there are holographic 
QCD models that are closer to a string-theoretic top-down construction such as the typically more involved models of Ref.~\cite{Casero:2007ae,Arean:2013tja,Giannuzzi:2021euy}. It would be interesting to see if they can achieve a better fit of masses and mixing angles of $f_1$, $f_1'$ mesons, and what they would predict for their contribution to $a_\upmu$.

\begin{table}[t]  
\bigskip
    \caption{Summary of the results \cite{Leutgeb:2022lqw} for the different contributions to $a_\upmu$ in the massive HW model due to Katz and Schwartz (but augmented by a nonzero gluon condensate (v1)) in comparison with the White Paper \cite{Aoyama:2020ynm} values. The mass of $\pi^0$ is fixed as input, the masses of $\eta$ and $\eta'$ are matched within 0.8 and 2.4\%, while the axial vector mesons $a_1$ and $f_1$ are between 4\% and 15\% too heavy, $f_1'$ by 28\%.}
    \label{tab:total}
\centering
\begin{tabular}{lccc}
\hline
$a_\upmu^{...}\times 10^{11}$ & v1(OPE fit) &  v1($F_\rho$-fit) & WP\phantom{$^{1^{1^1}}$} \\
 \hline
 $\pi^0$\phantom{$^{1^{1^1}}$} & 66.1 &  63.4 & 62.6$^{+3.0}_{-2.5}$ \\
 $\eta$ & 19.3 &  17.6 & 16.3(1.4) \\
 $\eta'$ & 16.9 &  14.9 & 14.5(1.9) \\
 $G/\eta''$ & 0.2 &  0.2 \\
 $\sum_{PS^*}$ & 1.6 & 1.4 \\[4pt]
 \hline
 PS poles\phantom{$^{1^{1^1}}$} & 104 &  97.5 & 93.8(4.0) \\
 \hline
 $a_1$\phantom{$^{1^{1^1}}$} & 7.8 &  7.1 \\
 $f_1+f_1'$  & 20.0 &  17.9 \\
 $\sum_{a_1^*}$ & 2.2 &  2.4 \\
 $\sum_{f_1^{(')*}} $ & 3.6 &  3.0 \\[4pt]
 \hline
 AV+LSDC\phantom{$^{1^{1^1}}$} & 33.7 &  30.5 & 21(16) \\
 \hline
 total\phantom{$^{1^{1^1}}$} & 138 & 128 & 115(16.5) \\
 \hline
\end{tabular}
\end{table}

\subsection*{Acknowledgments}
We would like to thank Luigi Cappiello,
Gilberto Colangelo, Giancarlo D'Ambrosio, Martin Hoferichter, Elias Kiritsis, and Pablo Sanchez-Puertas for useful discussions.
J.~L.\ and J.~M.\ have been supported by the Austrian Science Fund FWF, project no. P33655,
and by the FWF doctoral program
Particles \& Interactions, project no. W1252-N27.

%
\bibliography{hlbl}

\end{document}